\documentclass[12pt,preprint]{aastex}
\usepackage{epsfig}
\begin{document}

\title{Constraints on the acceleration of the solar system \\
from high-precision timing}

\shorttitle{Acceleration of the solar system}

\author{
Nadia L. Zakamska\altaffilmark{1}, 
Scott Tremaine\altaffilmark{1}}
\altaffiltext{1}{Princeton University Observatory, Princeton, New Jersey 08544}

\begin{abstract}
Many astronomers have speculated that the solar system contains undiscovered massive planets or a distant stellar companion. The acceleration of the solar system barycenter can constrain the mass and position of the putative companion. In this paper we use the most recent timing data on accurate astronomical clocks (millisecond pulsars, pulsars in binary systems and pulsating white dwarfs) to constrain this acceleration. No evidence for non-zero acceleration has been found; the typical sensitivity achieved by our method is $a_{\odot}/c\sim$ a few$\times 10^{-19}$ s$^{-1}$, comparable to the acceleration due to a Jupiter-mass planet at 200 AU. The acceleration method is limited by the uncertainties in the distances and by the timing precision for pulsars in binary systems, and by the intrinsic distribution of the period derivatives for millisecond pulsars. Timing data provide stronger constraints than residuals in the motions of comets or planets if the distance to the companion exceeds a few hundred AU. The acceleration method is also more sensitive to the presence of a distant companion ($\ga 300-400$ AU) than existing optical and infrared surveys. We outline the differences between the effects of the peculiar acceleration of the solar system and the background of gravitational waves on high-precision timing.
\end{abstract}


\section{Introduction}

For over 150 years, astronomers have wondered whether the solar system contains undiscovered massive planets or a distant stellar companion. Small residuals in the orbit of Uranus motivated the search for high parallax objects near the ecliptic that resulted in the discovery of Pluto in 1930 \citep{tomb61}. Numerous trans-Neptunian bodies have been discovered since, e.g., using the Caltech Wide Area Sky Survey \citep{truj01}, but all these objects (including Pluto) are too light to affect the major planets' orbits at the current measurement precision. \citet{stan93} argued persuasively that the claimed residuals in the orbit of Uranus were likely due to systematic errors, such as an underestimate of Neptune's mass by 0.5\%, rather than to an unseen massive object. The various constraints on the mass and position of a putative Planet X or a distant stellar companion were summarized by \citet{trem90} and \citet{hogg91}, who concluded that there was no dynamical evidence for an unseen companion down to the observational precision at the time of writing.

A massive companion would exert an acceleration on the known solar system barycenter. Such acceleration affects the observed value of the rate of the period change of astronomical clocks, such as pulsars and pulsating white dwarfs. In this paper we use the most up-to-date timing data to look for evidence of this acceleration. 

We consider a clock far outside the known planets' orbits, at distance $d$ in the direction given by unit vector $\bf n$ and moving with velocity $\bf v$ (all values are relative to the barycenter of the known solar system). Due to some physical process, the clock has an intrinsic period $P$ which changes at a rate $\dot{P}$. The observed period and period change rate are:
\begin{eqnarray}
P'=P\left(1+\frac{\bf n \cdot v}{c}\right)+O\left(\frac{v^2}{c^2}\right)\label{pprime}\\
\dot{P}'=\dot{P}\left(1+\frac{\bf n \cdot v}{c}\right) + P\left(\frac{v_T^2}{dc}+\frac{\bf n\cdot a}{c}\right)+O\left(\frac{v^2}{c^2}\right),\label{pdotprime}
\end{eqnarray}
where $v_T$ is the transverse velocity ($v_T=\mu d$, where $\mu$ is the proper motion) and $\bf a$ is the acceleration of the clock relative to the solar system barycenter. The component of this acceleration due to the solar system companion is denoted $- \bf a_{\odot}$ throughout this paper (so that the acceleration of the solar system due to the companion is $\bf a_{\odot}$). The neglected terms include the special relativistic time dilation which is of order $v^2/c^2$. In equation (\ref{pdotprime}) the intrinsic period $P$ can be replaced by the observed period $P'$ to the same order. 

A planet of mass $m_p$ at distance $d_p$ would exert an acceleration of 
\begin{equation}
\frac{a_{\odot}}{c}=\frac{G m_p}{c \, d_p^2}=2\times 10^{-18} \mbox{ s}^{-1} \left(\frac{m_p}{M_J}\right)\left(\frac{d_p}{100\mbox{ AU}}\right)^{-2},
\end{equation}
where $M_J$ is the mass of Jupiter; we consider timing of objects that can provide similar or better constraints. In Section \ref{sec_stat} we consider constraints on the acceleration in a statistical sense from timing of large groups of objects. In Section \ref{sec_ind} we consider acceleration constraints from individual objects. In Section \ref{sec_waves} we briefly discuss effects of a possible background of low-frequency gravitational waves on timing of astronomical clocks. We summarize our results in Section \ref{sec_conc}. 

\section{Statistical constraints on the acceleration}
\label{sec_stat}

\subsection{Statistical approach}

Pulsars are among the most stable astronomical clocks, their periodic signals being due to the rotation of a compact neutron star. About 1500 pulsars are known. Typically, the periods of pulsars slowly increase with time as they lose their rotational energy to particles and radiation. Whenever a secular period decrease is observed, it is believed that the observed value of $\dot{P}'$ is not intrinsic but rather due to the acceleration of the pulsar relative to the solar system. In support of this view, negative period derivatives have so far only been observed in globular clusters \citep{frei01}, arising from the pulsars' accelerations in the clusters' potentials. In most cases, extracting the acceleration using equation (\ref{pdotprime}) is problematic, since (i) the intrinsic period derivative is not known (first term of eq. \ref{pdotprime} on the right-hand side); (ii) line-of-sight velocities of pulsars have not been measured (second term); (iii) distances to pulsars are highly uncertain \citep{tayl93,bris02}; and (iv) proper motions have been reliably measured only for relatively nearby objects.  

We now proceed to estimating typical contributions of different terms to the observed period change rate in equation (\ref{pdotprime}). If we assume for a moment that the relative acceleration is at the level of $a/c\sim 10^{-18}$ s$^{-1}$, then for ordinary radio pulsars ($P\sim1$ s, $\dot{P}\sim2\times 10^{-15}$, $v\sim100$ km s$^{-1}$, $r\sim1$ kpc) the typical values are
\begin{equation}
\dot{P}'/\dot{P}\sim1+3\times 10^{-4}+5\times 10^{-4} + 5\times 10^{-4}\label{pdotp_psr}
\end{equation}
(in the same order as in eq. \ref{pdotprime}). The observed period change is dominated by the intrinsic value, and all other terms are tiny corrections of similar magnitudes. Thus ordinary radio pulsars are not useful for constraining the solar system acceleration on the order $a_{\odot}/c\la 10^{-18}$ s$^{-1}$. An additional complication is that normal pulsars concentrate strongly toward the Galactic plane and therefore cannot be used to constrain the component of the acceleration perpendicular to the plane. 

The period distribution of radio pulsars is bimodal \citep{phin94}, with a minimum at about 25 ms; there are only a few tens of objects with shorter periods. These millisecond pulsars (MSPs) typically have $P\sim10^{-2}$ s, $\dot{P}\sim4\times 10^{-20}$, $v\sim100$ km s$^{-1}$, and $r\sim1$ kpc. In this case,
\begin{equation}
\dot{P}'/\dot{P}\sim1+3\times 10^{-4}+3\times 10^{-1}+2.5\times 10^{-1}\label{pdotp_msp}.
\end{equation}
The proper motion and the possible acceleration-related corrections are comparable to the intrinsic value of the period derivative, whereas the radial velocity term can be neglected. 

In Section \ref{sec_msps} we implement a statistical approach to determining the solar system barycenter acceleration using MSPs. This approach is based on the plausible assumption that the intrinsic $\dot{P}$ distribution is independent of the position on the sky. Therefore, if the solar system accelerates relative to the MSP population, there would be a systematic dependence of the observed $\dot{P}'$ on the position of the pulsars in the Galaxy (this idea was first explored by \citealt{harr77} for a small sample of ordinary pulsars). Such a change can be detected, provided that it is not too small compared to the dispersion of values $\dot{P}'$ in the population. Therefore, in contrast to ordinary pulsars, MSPs offer the hope of measuring a solar acceleration.

\subsection{Data handling and limitations}
\label{sec_data}

\begin{figure}
\epsscale{1.0}
\plotone{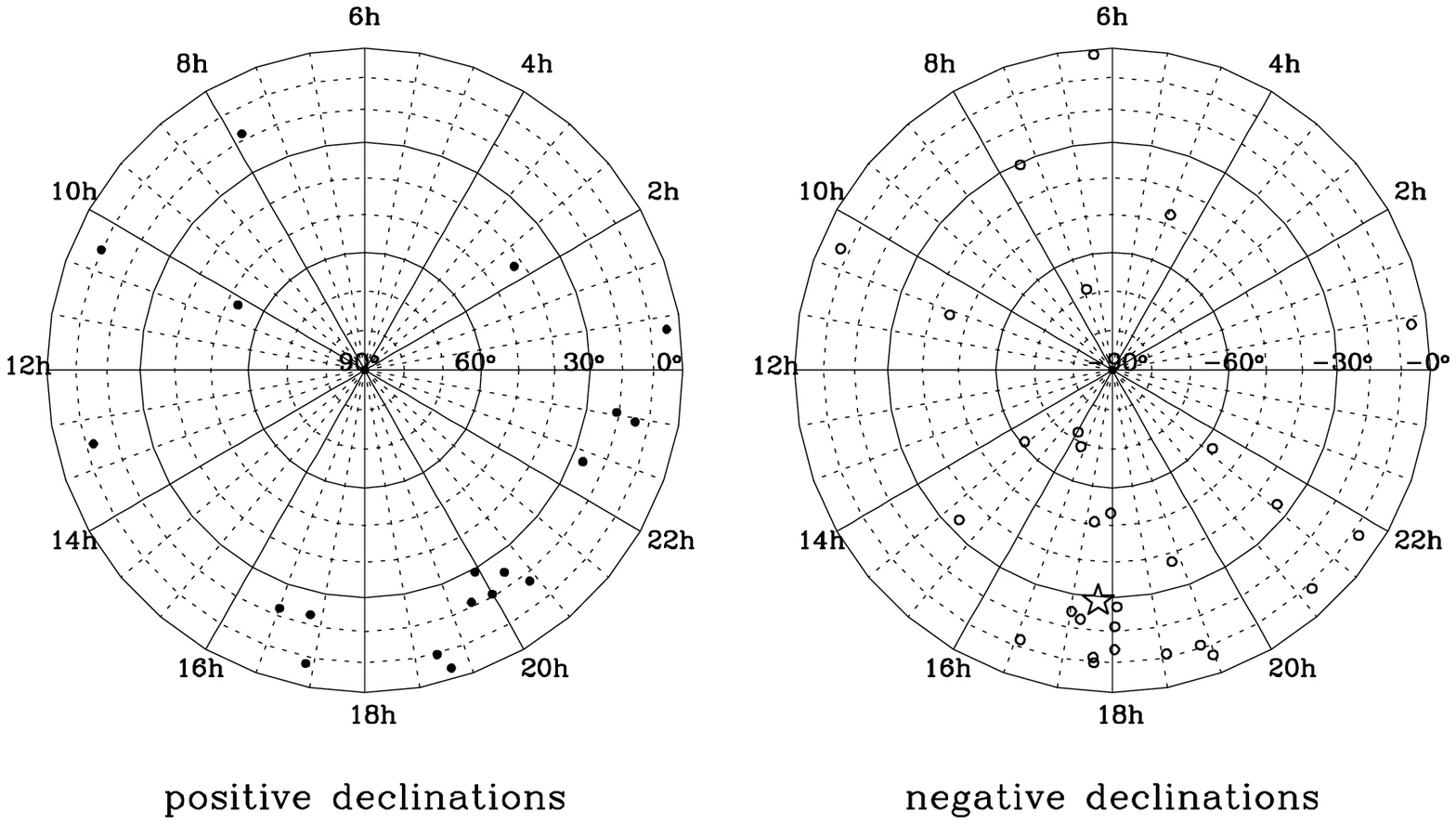}
\figcaption{Sky distribution of MSPs used for the analysis. 48 MSPs are marked with circles (filled for positive declinations and empty for negative declinations). For reference, the position of the Galactic center is indicated by a star. The same coordinate grid is used in all subsequent sky maps; the circles are lines of constant declinations, and right ascensions increase counter-clockwise. Equal solid angles are mapped into equal areas on this projection. \label{pic_objects}}
\end{figure}

From the ATNF pulsar database\footnote{Australia Telescope National Facility pulsar database at http://www.atnf.csiro.au/research/pulsar/psrcat/} \citep{manc05} as of January 2005, we selected all objects with $P'<25$ ms. We rejected all objects associated with globular clusters (as per \citealt{harr96}) since their accelerations relative to the solar system are likely to be dominated by the potentials of the systems they are residing in. We also neglected the young pulsar PSR~J0537$-$6910 from the Large Magellanic Cloud. We included only objects for which $\dot{P}'$ has been measured, producing a sample of 48 objects. Their distribution on the sky is shown in Figure \ref{pic_objects} which also introduces the sky visualization used throughout this paper. 

Many of the parameters given in the database such as the period rate changes, proper motions, parallaxes and orbital parameters (for those MSPs that are in binary systems) are obtained using the analysis of the times of arrival of the pulses, as described by \citet{kram04}. Of the 48 MSPs that we use, 35 are members of binary systems, and the period derivatives that we use have been corrected for the orbital motion. The relative errors of the period derivative measurements are typically better than 0.1\%, even for MSPs in binary systems, so they are a negligible source of uncertainty. Only eight MSPs in our sample have measured parallaxes, with a typical accuracy 30\%. Therefore, for most of the objects the distances were estimated using the measured values of the dispersion measure and the model of the distribution of the free electrons in the Galaxy by \citet{tayl93}. These estimates can be uncertain by up to 40\% \citep{bris02}; we discuss the effects of these uncertainties in Section \ref{sec_msps}. 28 out of 48 MSPs have proper motions measured as part of the timing analysis. 

All values listed in the database are corrected for the effects of the major bodies of the solar system and are given relative to the known solar system barycenter. Therefore, the accuracy of the solar system ephemerides limits the accuracy of the period derivative (and therefore of the relative acceleration). \citet{fair89} and \citet{damo91} estimate this accuracy to be $\sigma_{eph}/P=1.2\times 10^{-20}$ s$^{-1}$, where $\sigma_{eph}$ is the accuracy of the period derivative. The relative acceleration between a pulsar and the solar system that enters equation (\ref{pdotprime}) is composed of their respective accelerations in the Galactic potential and their peculiar accelerations due to interactions with neighboring objects. Were there a companion to the solar system, the acceleration $a_{\odot}$ it exerts must be greater than the peculiar acceleration of the solar system due to field objects in order to be detectable. The peculiar accelerations due to the field stars and giant molecular clouds in the solar neighborhood do not exceed $\sigma_{field}/P<1\times 10^{-20}$ s$^{-1}$ \citep{damo91}. We explicitly take into account the relative acceleration due to the overall Galactic potential as described in Section \ref{sec_msps}. 

Finally, a companion to the solar system would be displaced as a result of its orbital motion on the time scale that the pulsar timing exists (30$+$ years, or 20 years in the case of MSPs) leading to a change of the direction of the acceleration exerted on the solar system. This change is quite significant at the distance of Pluto, with angular motion of $30\deg\times \left(d_p/40\mbox{ AU}\right)^{-3/2}\left(t/20\mbox{ years}\right)$, but becomes negligible for distances to the companion $d_p\ga 100$ AU, or if we primarily use more recently discovered MSPs. Assuming one of these latter conditions holds, we neglect the change of the direction of the acceleration in all subsequent discussion.

\subsection{Millisecond Pulsars}
\label{sec_msps}

For each direction on the sky ($\alpha,\delta$) we construct two datasets based on the observational data for the 48 MSPs described above (in practice we used a grid spaced by 10 degrees in right ascension and declination, as shown in Figure \ref{pic_objects}). The first dataset contains $\dot{P}_i'/P_i'$ for all MSPs and the second dataset consists of $\cos\theta_i$, where $\theta_i$ is the angle between ($\alpha,\delta$) and the direction to the object ($\alpha_i,\delta_i$). If the solar system accelerates with $a_{\odot}$ in the direction ($\alpha, \delta$), from equation (\ref{pdotprime}) we know that for all objects there is a term $-a_{\odot} \cos\theta/c$ which contributes to the observed $\dot{P}'/P'$ value. If there is no acceleration in this direction, there should be no correlation between the observed timing properties of MSPs and their position on the sky (the null hypothesis). 

The presence of the correlation is assessed using the Spearman rank correlation coefficient which is calculated after replacing all values in a dataset by their ranks (e.g., the minimum value is replaced by 1 and the maximum by $N$ if the dataset consists of $N$ values). The advantage of using the rank sets is that the results do not depend on the distribution from which the datasets were drawn \citep{pres92}. The rank correlation coefficient $r_s$ is a signed quantity (positive if one value on average increases as a function of the other and negative if it decreases); values of $r_s$ around 0 imply there is no correlation, whereas values close to 1 or $-1$ indicate a strong correlation. The significance of the rank correlation coefficient can be assessed using the probability $\Pi(r_s)$ that a given value of $r_s$ is obtained for two uncorrelated sets of data.  

\begin{figure}
\epsscale{1.0}
\plotone{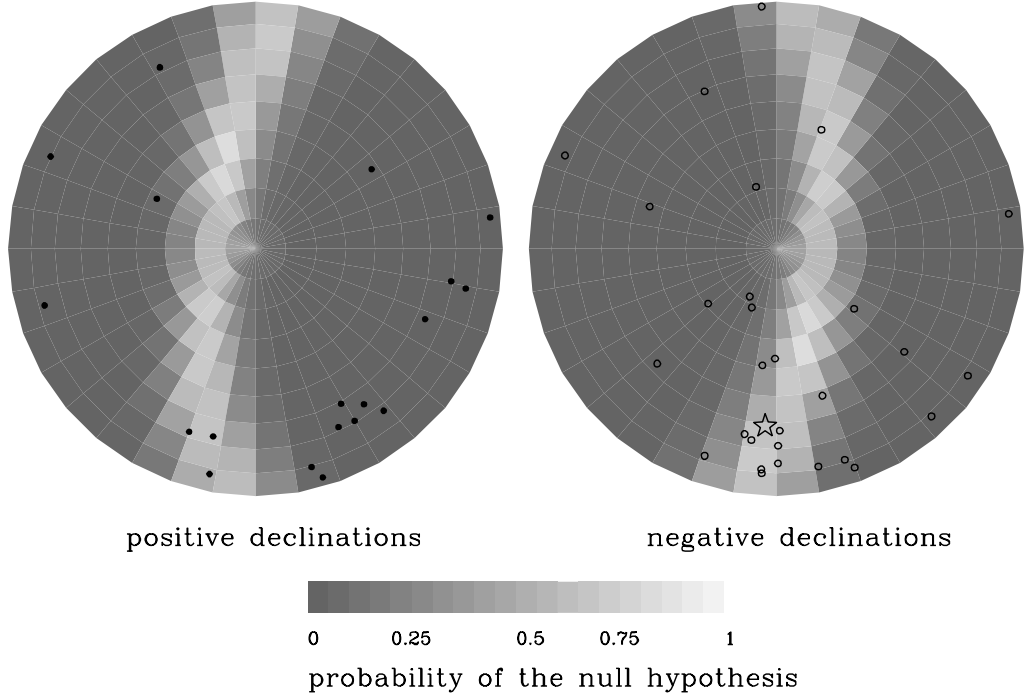}
\figcaption{The probability map for a simulated population of pulsars with the same positions as observed MSPs, but with $P=0.007$ s and $\dot{P}=2.4\times 10^{-20}$, if the solar system accelerates relative to this population with $a_{\odot}/c=10^{-18}$ s$^{-1}$ in the direction $\alpha_{\odot}=180$, $\delta_{\odot}=0$. Each area on the sky around the direction ($\alpha,\delta$) is shaded according to the probability that $\dot{P}_i'/P_i'$ is not correlated with $\cos\theta_i$ relative to this direction. The darker the shading, the smaller the probability of the null hypothesis (no correlation), and the stronger is the correlation in this direction. The acceleration component is clearly visible (the light-colored band is the plane of the least correlation, which lies perpendicular to the direction of the acceleration); this highly significant detection of an acceleration has $\langle\Pi\rangle=0.113$, min$(\Pi)\simeq 0$, max$(\Pi)\simeq 1$. The acceleration that can be retrieved based on the probability map is slightly offset relative to the input acceleration; this is due to the inhomogeneous distribution of objects on the sky. By construction, probability maps are invariant with respect to reversing the direction of the acceleration, $\Pi(\alpha,\delta)=\Pi(\pi+\alpha,-\delta)$. The coordinate system and the legend are the same as in Figure \ref{pic_objects}. \label{pic_test_18}}
\end{figure}

For each direction we evaluate the rank correlation coefficient $r_s(\alpha,\delta)$ between the two datasets $\dot{P}_i'/P_i'$ and $\cos\theta_i$. We then evaluate the probability of the null hypothesis $\Pi(\alpha,\delta)$ (i.e., the probability that $\dot{P}_i'/P_i'$ and $\cos\theta_i$ are uncorrelated) and use these values to construct a ``probability map''. An example of a probability map is given in Figure \ref{pic_test_18} for a simulated dataset. For this simulation we used the real positions of MSPs, but assigned them all the same $P$ and $\dot{P}$ values and added an artificial acceleration to the data producing slightly different observed $\dot{P}'$ values. The probability map for the negative declinations is fully determined by the probability map for positive declinations, since the calculated probabilities by construction are invariant with respect to reversing the direction, $\Pi(\alpha,\delta)=\Pi(\pi+\alpha,-\delta)$.  

\begin{figure}
\epsscale{1.0}
\plotone{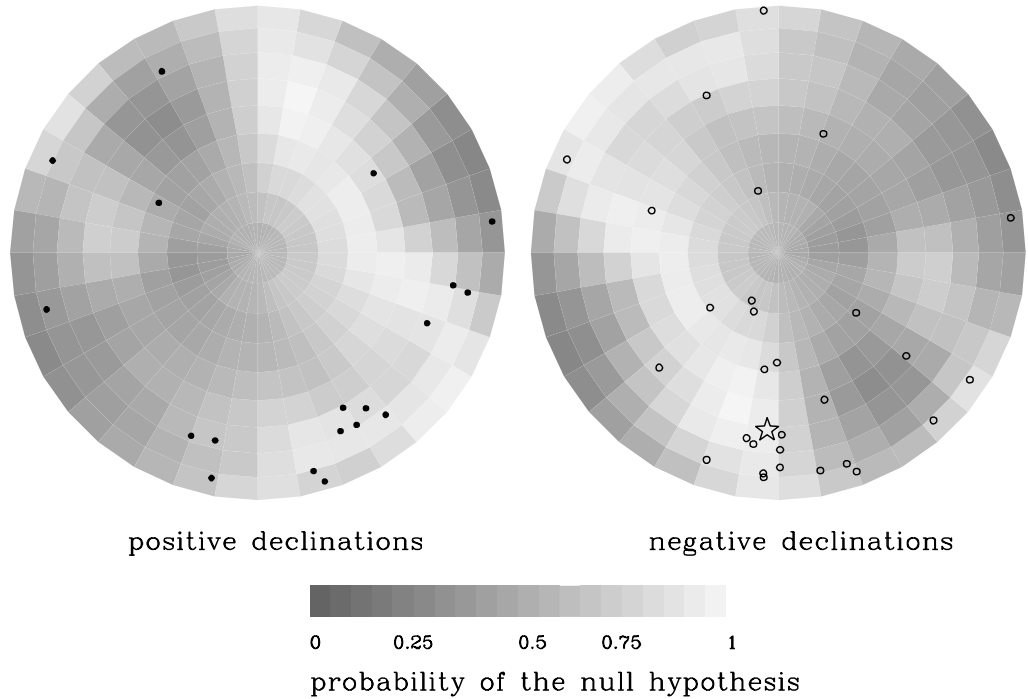}
\figcaption{The probability map for the real data on MSPs corrected for the proper motion effect and the relative accelerations in the Galactic potential field. On this plot max$(\Pi)=0.996$, min$(\Pi)=0.213$, $\langle\Pi\rangle=0.610$. No acceleration signature is present in this map, as indicated by the high values of $\Pi$ (lighter shading compared to the previous Figure) and by the lack of reflection symmetry about a plane. The coordinate system and the legend are the same as in Figure \ref{pic_objects}. \label{pic_real}}
\end{figure}

To produce the probability map using the real data on MSPs we corrected the observed period derivatives for the proper motion effect and the relative acceleration in the Galactic potential using equation (\ref{pdotprime}). To this end, we constructed a new dataset consisting of $\dot{P}_{i,\,new}/P_i'\equiv\dot{P}_i'/P_i'-\mu_i^2 d_i/c-a_{gal, i}/c$, where $\mu_i$ are the observed proper motions of the MSPs (consistent with zero for 20 out of 48 objects), $d_i$ are their distances from the sun and $a_{gal, i}$ are the accelerations of the pulsars relative to the solar system in the Galactic potential projected onto the direction to the pulsars. The typical values of the proper motion correction are
\begin{equation}
\frac{\dot{P}_{i,\,pm}}{P}=\frac{\mu_i^2 d_i}{c}=2.5\times 10^{-19}\mbox{ s}^{-1}\,\left(\frac{\mu_i}{10\mbox{ mas year}^{-1}}\right)^2\left(\frac{d_i}{1\mbox{ kpc}}\right).
\end{equation}
The correction for the relative Galactic acceleration is performed using the Galactic potential in the form suggested by \citet{pacz90} (hereafter `P90 model') which consists of a disk, a spheroid (bulge) and a spherically symmetric halo; the model is normalized to have the rotation velocity of 220 km s$^{-1}$ at 8 kpc. This model of the Galactic potential was updated by \citet{dehn98}. Numerically, the P90 and the \citet{dehn98} models produce similar results \citep{sun04}, but the P90 potential is easier to use since it has an analytic form. The typical values of this correction are a few times smaller than the acceleration of the Sun toward the center of the Galaxy, $a_{gal}/c \sim a_{gal,\odot}/c \times (d_i/8\mbox{ kpc})$, where $a_{gal,\odot}/c=6.5\times 10^{-19}$ s$^{-1}$. The probability map corrected for proper motions and for relative Galactic accelerations is shown in Figure \ref{pic_real}. 

In contrast to the probability map for the simulated accelerated dataset (Figure \ref{pic_test_18}), the probability map based on the real data (Figure \ref{pic_real}) does not show strong correlations in any direction and has on average much higher probabilities $\Pi$ of the null hypothesis. The lowest value of $\Pi$ in any direction in Figure \ref{pic_real} is 0.213, which is consistent with the null hypothesis. Therefore, there is no obvious evidence for a non-zero acceleration of the solar system relative to the MSP population. 

\begin{figure}
\epsscale{1.0}
\plotone{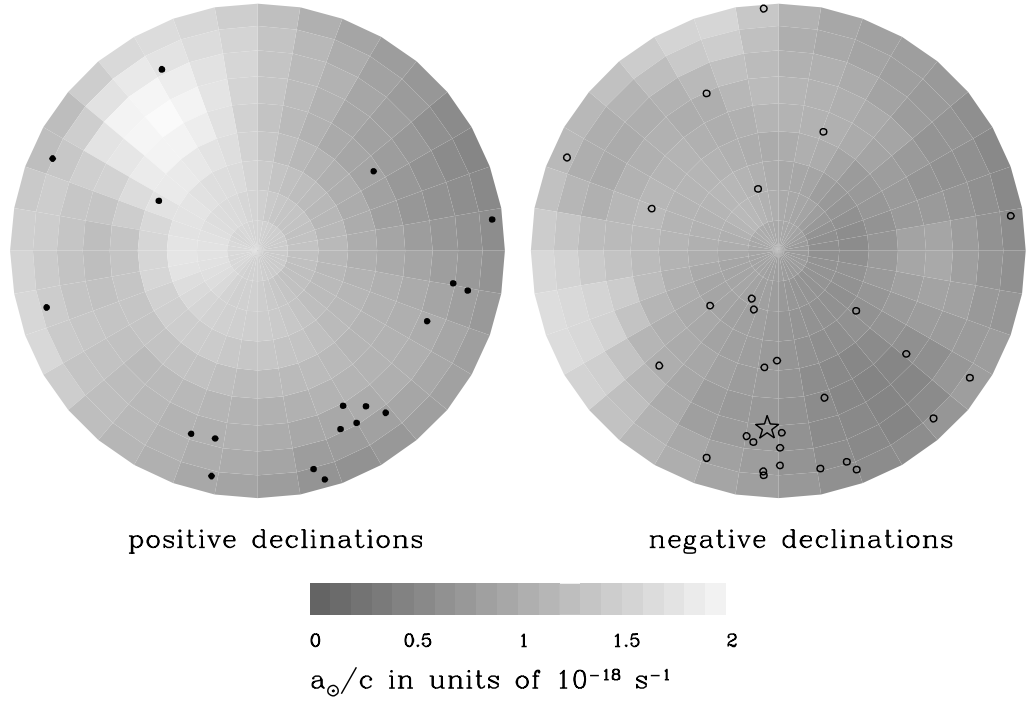}
\figcaption{Values of the acceleration that can be ruled out at a 95\% confidence level in each direction based on timing of MSPs. The data have been corrected for proper motions and Galactic accelerations. The coordinate system and the legend are the same as in Figure \ref{pic_objects}. \label{pic_confidence}}
\end{figure}

To establish the significance of this null result and to find the maximum acceleration allowed by the data on MSPs, we first created 1000 made-up datasets by using the positions of the real objects and randomly shuffling their $\dot{P}_{new}'/P'$ values. For each direction on the sky ($\alpha,\delta$), these shuffled datasets provide a distribution of rank correlation coefficients of uncorrelated datasets. We then correct the dataset $\dot{P}_{i,new}'/P_i'$ for a putative acceleration $a_{\odot}$ directed toward ($\alpha,\delta$) and calculate the rank correlation coefficient for this corrected dataset. We keep increasing the value of $a_{\odot}$ until the data start showing the presence of a correlation in a statistical sense, i.e., when the rank correlation coefficient reaches its top 5\% value as determined from the shuffled datasets (which are presumed uncorrelated with the position on the sky). When such value of $a_{\odot}$ is reached, we have overcorrected for the acceleration, and this value is ruled out by the data at the 95\% level. We calculate these limiting values of acceleration for each direction on the sky and show them in Figure \ref{pic_confidence}. 44\% of the sky has limiting accelerations $a_{\odot}/c<10^{-18}$ s$^{-1}$, 99\% of the sky has limiting accelerations $a_{\odot}/c<2\times 10^{-18}$ s$^{-1}$ and 100\% of the sky has limiting accelerations $a_{\odot}/c<3\times 10^{-18}$ s$^{-1}$. 

None of the conclusions of this Section change if we use the observed values of $\dot{P}'/P'$ without correcting for the proper motion effects and the relative Galactic accelerations. The corresponding probability map appears different in detail from the one shown in Figure \ref{pic_real}, but it has a similar value of $\langle\Pi\rangle$ and $\min(\Pi)$ and shows no sign of a correlation. The limiting accelerations are also similar to the values found in the preceding paragraph. This finding is consistent with the fact that the limiting accelerations are at least a few times larger than the typical values of these corrections. This also means that the uncertainties in the distances to MSPs, however large, do not seriously affect the statistical method. The method is ultimately limited by the relatively broad distribution of the intrinsic values of $\dot{P}/P$, as illustrated by equation (\ref{pdotp_msp}).

\section{Individual objects}
\label{sec_ind}

Unlike the case of MSPs, where the intrinsic values of $\dot{P}/P$ are unknown, in some objects these values are known from theoretical arguments. Since the relative acceleration is constrained by the difference between the observed and the theoretical period change rates $(\dot{P}_{obs}-\dot{P}_{th})/P$, both $\dot{P}_{obs}$ and $\dot{P}_{th}$ must be measured or constrained with enough precision to produce an interesting estimate of acceleration. Therefore, we consider only objects for which $\dot{P}_{obs}/P\la 10^{-17}$ s$^{-1}$ or so. 

The first group of such objects is short-period binary systems containing a pulsar (see \citealt{stai04} for a recent review). From this list we excluded all objects found in globular clusters, which might be affected by peculiar accelerations associated with the cluster potential. Of the remaining binary systems there are five in which the derivative $\dot{P}_b$ of the orbital period $P_b$ has been measured (Table 1), including the most recent measurement for the binary pulsar PSR~J0737$-$3039 \citep{burg03, lyne04, kram05}. In addition, there are five more binary systems for which interesting observational constraints exist on the orbital period decay. These ten systems are listed in Table 1 grouped by the accuracy of the constraint that they provide (the six objects on top of the Table form the higher accuracy `group 1'). In these objects, the intrinsic value of $\dot{P}_b$ can be calculated under the assumptions that the orbital period decay is due to emission of gravitational waves and that this emission is correctly described by general relativity.

The second group of objects with potentially known intrinsic $\dot{P}/P$ is pulsating white dwarfs. In this case $P$ refers to the period of any of the oscillation modes. Pulsating white dwarfs come in three types, based on their effective temperature \citep{muka03}; nearly a hundred objects are known (SIMBAD\footnote{SIMBAD astronomical database, http://simbad.u-strasbg.fr}; \citealt{muka04}). While the typical periods of the oscillation modes are a few hundred seconds in all three classes, the period derivatives range from $10^{-11}$ to $10^{-15}$, decreasing for cooler objects \citep{sull98}. Because of our precision requirements, only the coolest pulsating white dwarfs (typically spectroscopically classified as hydrogen-rich, or DA) are suitable for acceleration constraints. In these stars the intrinsic value $\dot{P}$ is determined by the rate of cooling, while other factors such as gravitational contraction are unimportant \citep{kepl00}. Cooling of DA white dwarfs has been modeled leading to estimates of the intrinsic $\dot{P}/P$ \citep{brad98}. However, the period derivative has been measured only for a DA white dwarf G117$-$B15A, which is the only one that we include in our analysis \citep{kepl00}. For some other objects (e.g., the prototype of the class ZZ Ceti, \citealt{muka03}) upper limits on $\dot{P}_{obs}$ exist, but they are less restrictive than the actual measurement for G117$-$B15A.

Kinematic corrections due to proper motions $\dot{P}_{pm}\pm \sigma_{pm}$ and due to relative Galactic accelerations $\dot{P}_{gal}\pm \sigma_{gal}$ need to be taken into account for all objects that we use, in order to meaningfully compare the observed $\dot{P}_{obs}\pm \sigma_{obs}$ and the theoretical $\dot{P}_{th}\pm \sigma_{th}$ values. Timing studies listed above and in Table 1 typically include estimates of these corrections, unless the authors believe the corrections are unimportant. In one case (PSR~B1913+16, \citealt{damo91}) the kinematic correction also includes accelerations induced on the sun and the pulsar by their respective neighbor stars (with typical distances $\ga 1$ pc) and giant molecular clouds, but these authors find that these accelerations are tiny compared to the Galactic acceleration and the proper motion effect. 

We calculated the Galactic acceleration corrections using the Milky Way potential in the form suggested by P90. The errors of these corrections were calculated including both the uncertainty in the distance and the uncertainty of the model of the potential. The uncertainty in the model can be estimated by comparing the P90 potential to other suggested forms of potential, for example, the most recent update by \citet{dehn98}. For each object, the relative error in the acceleration component parallel to the Galactic plane is calculated by adding in quadrature the relative error in the distance between the sun and this object and 15\%. This last value comprises the uncertainties of the potential model in the plane ($\la 10\%$, \citealt{sun04}), the uncertainty in the solar Galactic velocity ($\la 10\%$) and the uncertainty in the distance to the Galactic center ($\la 10\%$). The relative error in the acceleration component perpendicular to the plane is obtained by adding in quadrature the relative distance error and 25\% which is dominated by the uncertainty in the potential model \citep{sun04}.

From equation (\ref{pdotprime}), the acceleration of the Solar system in any direction can be constrained from each object with a known $\dot{P}_{obs}$ and $\dot{P}_{th}$:
\begin{equation}
-P \,\frac{\bf a_{\odot} \cdot n}{c}=\dot{P}_{obs}-\dot{P}_{th}-\dot{P}_{pm}-\dot{P}_{gal}\pm\sqrt{\sigma_{obs}^2+\sigma_{th}^2+\sigma_{pm}^2+\sigma_{gal}^2}\equiv \Delta\dot{P}\pm\sigma_{tot}, \label{eq_a_pbdot}
\end{equation} 
where the unit vector $\bf n$ points toward the object that we are using. If only one object is available, the acceleration is unconstrained in the plane perpendicular to the line of sight. Therefore, the constraint on the acceleration can be improved by combining data on several objects in different parts of the sky.

In Table 1 we list all the values that enter equation (\ref{eq_a_pbdot}) and their errors for the ten pulsars in binary systems and one DA white dwarf used in our analysis. The objects PSR~B1913+16 and PSR~J1713+0747 provide the most interesting constraints on the acceleration. For PSR~B1913+16, the accuracy improved by about an order of magnitude in the last 13 years (cf. \citealt{weis04} and \citealt{damo91}). This object has been timed for 30 years and has been used to constrain the acceleration of the solar system all by itself \citep{thor85}. In the object PSR~J1713+0747 the orbital period decay has not been detected, but the available upper limit on the orbital decay based on 12 years of observations combined with the long orbital period of this system (68 days) makes it useful for our analysis. 

\begin{figure}
\epsscale{1.0}
\plotone{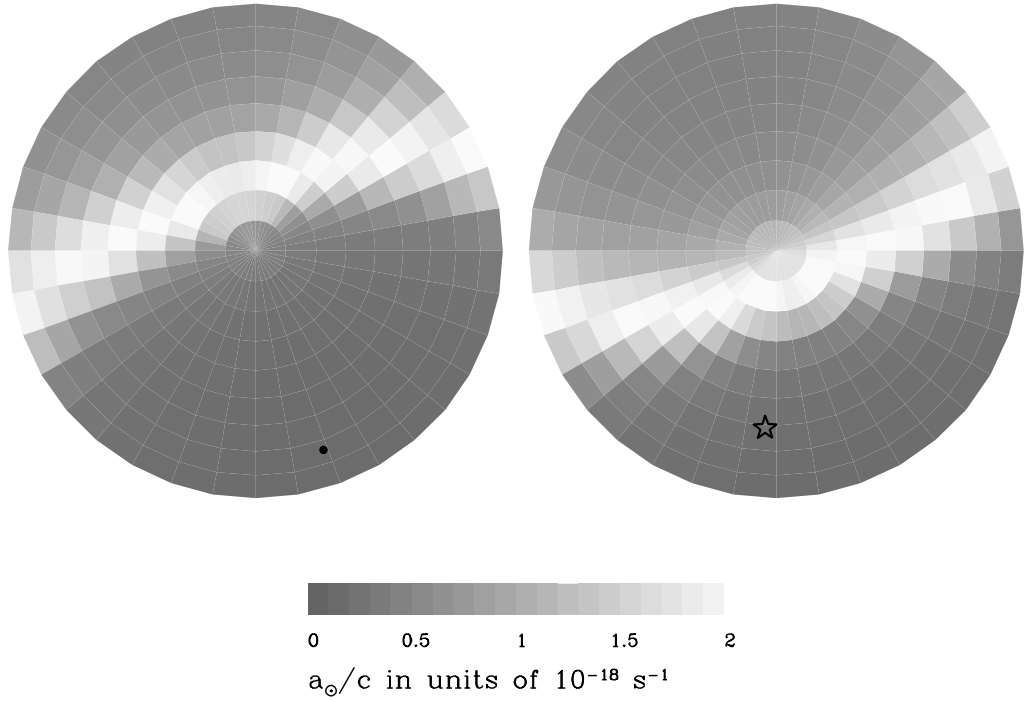}
\figcaption{Values of the acceleration of the solar system that can be ruled out at a 2$\sigma$ confidence level in each direction based on the observations of PSR~B1913+16 alone. The most poorly constrained directions are those perpendicular to the line of sight to the pulsar, hence the symmetric appearance. The coordinate system and the legend are the same as in Figure \ref{pic_objects}. \label{pic_a_B1913}}
\end{figure}

\begin{figure}
\epsscale{1.0}
\plotone{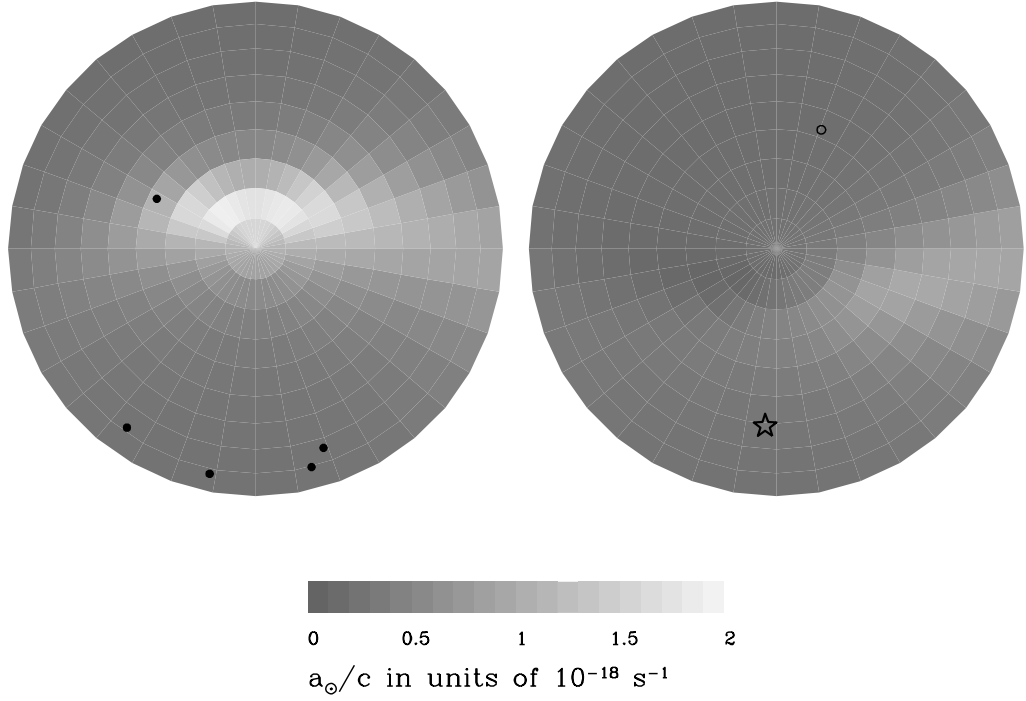}
\figcaption{Values of the acceleration that can be ruled out at a 2$\sigma$ confidence level in each direction based on the observations of six `group 1' objects from Table 1. Adding the remaining five objects has almost no effect on the acceleration constrains. The coordinate system and the legend are the same as in Figure \ref{pic_objects}. \label{pic_a_six}}
\end{figure}

\begin{figure}
\epsscale{1.0}
\plotone{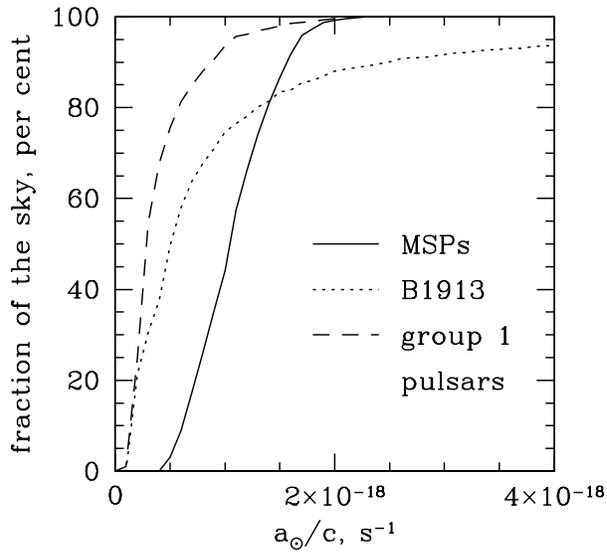}
\figcaption{The values of acceleration that can be ruled out at 95\% level by different methods: statistical analysis using MSPs (solid line), timing of PSR~B1913+16 alone (dotted line), timing of the six `group 1' pulsars combined (dashed line). In the latter case, adding the remaining five objects in Table 1 has almost no effect on the acceleration constraints. This Figure is the graphical complement to Table 2. \label{pic_sum}}
\end{figure}

The best-fit acceleration in the direction $(\alpha,\delta)$ from object $i$ is given by $a_{\odot}/c=-(\Delta \dot{P}_i \pm \sigma_{tot, i})/(P_i \cos\theta_i)$ where $\theta_i$ is the angle between $(\alpha,\delta)$ and the direction to the object. There is no evidence for a non-zero acceleration at the $2\sigma$ level for any object. In Figure \ref{pic_a_B1913} we show the values of accelerations ruled out at a 2$\sigma$ level for each direction on the sky based on the observations of PSR~B1913+16 alone. The least constrained directions are those perpendicular to the direction to the pulsar. The constraints in these regions are significantly improved when we include another five objects from the high-accuracy group, as shown in Figure \ref{pic_a_six}. In this Figure, for each position on the sky we calculated the error-weighted mean and the error-weighted variance of the accelerations obtained using equation (\ref{eq_a_pbdot}) for the six pulsars and plotted (mean + 2 $\times$ variance$^{1/2}$). Adding the remaining five objects from Table 1 has almost no effect on the acceleration constraints. Table 2 and Figure \ref{pic_sum} summarize what values of acceleration of the solar system can be ruled out by different methods.

The method described here assumes that the peculiar accelerations of each object due to nearby stars and clouds are negligible compared to their Galactic accelerations relative to the sun. All objects used in our calculation are either in the disk of the Galaxy or somewhat above it, rather than in the bulge, and most are within 1 kpc, so the estimate of the peculiar acceleration of the sun (Section \ref{sec_data}) is likely to apply to all these objects as well. If so, the peculiar acceleration is negligible compared to the uncertainties introduced by the distance determination.

\section{Constraints on the background of gravitational waves}
\label{sec_waves}

The universe may be filled with gravitational waves, produced either by individual sources such as supernovae or merging stars, or at very early stages of the expansion shortly after the Big Bang \citep{alle97}. A passing wave affects the propagation of electromagnetic emission and produces a difference between the observed and the emitted frequencies. Specifically, the frequency change due to gravity waves is \citep{mash79}
\begin{equation}
\frac{\nu_{obs}-\nu_{em}}{\nu_{em}}\sim \eta_{em}-\eta_{obs},
\label{eq_gwshift}
\end{equation}
where $\nu$ is the frequency and $\eta \ll 1$ is the perturbation of the metric due to gravitational waves at the position of the source ($\nu_{em}$, $\eta_{em}$) and of the observer ($\nu_{obs}$, $\eta_{obs}$). Similarly, the observed rotational period and period change of a pulsar are different from the intrinsic values in the presence of gravitational waves \citep{sazh78, detw79}. Of course, in order to use these differences to compute the effects of the gravitational waves, we now have to assume that there is no unknown contribution to the observed period change (i.e., from a massive companion). 

For example, gravitational waves with frequencies $f\ga 1/T$, where $T=10-30$ years is the span of pulsar timing, introduce noise to the timing data. With some assumptions about the spectral energy distribution of the waves, one can constrain the amplitude of the gravitational waves (and therefore their energy density) from observations of individual pulsars \citep{detw79, roma83, kasp94, thor96, mchu96} or pulsar pairs using cross-correlation of timing noise from two objects \citep{hell83}. This method is sensitive to frequencies $1/T\la f\la 1/\tau$, where $\tau$ is a typical spacing between consecutive measurements, and the most stringent available upper limit on the energy density in such waves, expressed as a fraction of the critical density of the universe, is $\Omega_{GW} h^2\la 1.0\times 10^{-8}$ \citep{thor96, mchu96}, where $H_0=h\times 100$ km s$^{-1}$ Mpc$^{-1}$ is the Hubble constant. 

In this paper we have used the values of the period and the period derivative which are the best-fit timing solutions over the entire span of available data, thereby discarding possible variations with frequencies $f\ga 1/T$. If we now focus on stochastic waves with frequencies $c/d\la f\la 1/T$ (where $d$ is the typical distance to the pulsars from the sun), then the $\eta_{em}$ terms of equation (\ref{eq_gwshift}) are uncorrelated for different pulsars, but the second term of this equation provides a common change of periods and period derivatives for all objects. The difference between the expected and the measured values of the period derivatives in individual objects, such as those listed in Table 1, can be directly used to constrain the energy density in gravitational waves in this frequency regime \citep{bert83, tayl89, thor96}, giving $\Omega_{GW} h^2 \la 0.04$ \citep{thor96}. The accuracy of the $\Delta \dot{P}/P$ measurement for PSR~B1913+16 (which dominated this constraint) has improved by about a factor of three since the publication by \citet{thor96}, so the current upper limit on $\Omega_{GW}h^2 \propto (\Delta \dot{P}/P)^2$ is approximately an order of magnitude better. In contrast to the case of the peculiar solar acceleration, the effect of gravitational waves in this frequency range on timing measurements does not depend on the position on the sky, so it cannot be extracted statistically from the timing data as we did in Section \ref{sec_stat} for the solar acceleration. 

Finally, for very low frequency gravitational waves ($f\la c/d$) the size of the `detector' (the solar system and the pulsars) is smaller than the wavelength of the waves, and it can be shown that, to first order in $\eta$, the change in the observed period of the signals is 
\begin{equation}
\frac{P'-P}{P}=\frac{d}{2c}\sum_{i,j} n_i n_j \dot{\eta}_{ij},
\label{eq_gwlong}
\end{equation} 
where $n_i$ are the spatial components of the unit vector in the direction to the pulsar and $\eta_{ij}$ are the spatial components of the metric perturbation taken at the observer's position. This regime can be probed both with individual objects (by taking a time derivative of eq. \ref{eq_gwlong} and using limits on $\Delta \dot{P}$ from Table 1) and in the statistical sense using MSPs since the effect is expected to increase with distance. In the latter case, one must be aware that a systematic dependence on the distance may be introduced when applying the proper motion correction to the observed values $\dot{P}'_i/P_i$, since $\mu_i$ are typically not known for more distant objects. For the subset of MSPs with measured proper motions, the correlation of $\dot{P}_i/P_i$ versus $d_i$ is present at less than 70\% confidence level (using the rank correlation test). For individual objects, the amplitude of the metric perturbation is best constrained with PSR~J1713+0747 and PSR~B1913+16 and is consistent with 0 within 1.1$\sigma$. Using equation (\ref{eq_gwlong}) and the expression for energy density of gravitational waves (e.g., \citealt{detw79}) we can place an upper limit on the energy density of very low frequency waves:
\begin{equation}
f\Omega_f \sim \left(\frac{\Delta \dot{P}}{P}\right)^2 \left(\frac{c}{d\,f}\right)^2 \frac{1}{H_0^2}, \quad\mbox{where } \frac{c}{d\,f}\gg 1.
\label{eq_fof}
\end{equation}
With the best accuracy of $\sigma_{tot}/P\simeq 1\times 10^{-19}$ s$^{-1}$ for objects in Table 1, equation (\ref{eq_fof}) implies $f\Omega_f h^2 \la 0.004\, (c/d\,f)^2$ (95\% confidence), where $\Omega_f {\rm d} f$ is the density relative to the critical density in the frequency range $(f,f+{\rm d}f)$. 

\section{Conclusions}
\label{sec_conc}

In the presence of an undiscovered massive companion, the known solar system barycenter would experience an acceleration which could be detected in observations of stable astronomical clocks. No such acceleration has been detected so far. The best limits on the acceleration are provided by the sources for which the intrinsic period change rate can be somehow estimated and then compared with the observed period change. Examples of such systems are pulsars in binary systems in which the orbital period decays due to emission of gravitational waves, and white dwarfs in which the pulsation periods change as the star cools. Alternatively, statistical limits on the acceleration can be provided by a group of objects of the same type distributed across the sky, even if the intrinsic period change rate of these objects is unknown. An example is millisecond pulsars, which provide sensitivity to the acceleration similar to that obtained using pulsars in binary systems and white dwarfs. The limits on the acceleration of the solar system barycenter that we obtained by different methods are summarized in Table 2.

The acceleration exerted by the companion on the solar system is $a_{\odot}/c=Gm_p/(c d_p^2)$, where $d_p$ is the current distance to the companion and $m_p$ is its mass. Therefore, the upper limit on the mass of the companion as a function of the upper limit on the acceleration is
\begin{equation}
m_p=0.26 M_J \times \left(\frac{a_{\odot}/c}{5\times 10^{-19}\mbox{ s}^{-1}}\right)\left(\frac{d_p}{100\mbox{ AU}}\right)^2.
\label{eq_conclusion}
\end{equation}
Companions with masses higher than this can be ruled out at a 2$\sigma$ confidence over about 76\% of the sky (Table 2). Our limit on the acceleration is about an order of magnitude better than the one available 15 years ago \citep{trem90} because of the improved accuracy of the timing observations of PSR~B1913+16 and because several other objects became available. 

Timing of pulsars has been widely used to place constraints on the energy density of low-frequency gravitational waves. In Section \ref{sec_waves} we summarized these methods and outlined the differences between the effects of a companion and of the gravitational waves on pulsar timing.

The observational accuracy will improve with time for many objects, orbital period decay will likely be measured for many pulsars in binary systems, and more millisecond pulsars will be discovered. For some objects used in our analysis further observations are unlikely to sharpen the limit on the acceleration since the current constraint that they provide is determined mostly by the uncertainties in the distances to pulsars and in models of the potential of the Galaxy. This limit has now been reached for PSR~B1913+16 and PSR~B1534+12 (Table 1), and there is little prospect for more accurate independent distance measurements \citep{weis04}. The limit on the acceleration is significantly tightened by adding objects which are closer and have a parallax measurement with better than 10\% precision, such as PSR~J1737+0747 (the acceleration constraint provided by this object is still dominated by the timing precision and will therefore improve with time). The statistical method using MSPs is currently limited by the intrinsic distribution of the period derivatives and is unlikely to rapidly improve until much larger samples of MSPs become available. Accelerations exerted by field objects and the uncertainties in solar system ephemerides are negligible in both methods.

Constraining the acceleration of the solar system using individual objects like pulsars in binary systems and white dwarfs relies on the theoretical arguments that are used to calculate the intrinsic period derivative. The reversed sequence has been used for some of these objects. For example, PSR~B1913+16 was used to confirm the prediction of general relativity for the rate of emission of gravitational waves assuming that the solar acceleration is negligible \citep{damo91, weis04}. Similarly, G117$-$B15A has been used to constrain the variation of the gravitational constant $\dot{G}$ \citep{benv04, bies04} and other exotic physics \citep{bies02}.

The acceleration of the solar system barycenter is not the only constraint on the mass and the distance of the companion; other methods have been described in detail by \citet{trem90} and \citet{hogg91}. One of the most sensitive of these is constraining the tidal force exerted by a distant companion ($\propto m_p/d_p^3$) using the ephemerides of known planets and comets. Depending on the assumed accuracy of the measured ephemerides, the masses that can be ruled out by the acceleration method and by the tidal method are similar for $d_p\simeq 400$ AU or so, but the acceleration method becomes more sensitive if the companion is at a larger distance (the limiting $m_p$ is $\propto d_p^2$ rather than $d_p^3$). Furthermore, the tidal method is much less sensitive away from the ecliptic plane, whereas the acceleration method can provide a uniform sensitivity over the entire sky. On the downside, the acceleration method is not sensitive to an extended symmetrical disk such as a massive Kuiper belt, whereas the tidal method can detect such structures, though with reduced sensitivity. 

The {\it Voyager} and {\it Pioneer} spacecraft have provided distances to Uranus and Neptune with unprecedented precision using ranging measurements during flybys \citep{ande89, ande95b}. However, the high accuracy of these measurements does not translate directly into a significant improvement in the ephemerides of the planets, since the ranging measurements constrain only the radial component of the position, whereas the transverse component is still determined using optical astrometric data. As a result, the constraints on the unseen mass in the solar system derived from the planet ephemerides that incorporate ranging data \citep{ande95b} are not very different from those derived from the optical ephemerides alone \citep{hogg91}. In addition to providing planet distances and masses, spacecraft themselves serve as accurate gravity probes. The two {\it Pioneers} are especially suitable \citep{ande02} as they did not require many reorientation maneuvers (which complicate modeling of trajectories). An explicit search for Planet X in the {\it Pioneer} tracking data yielded a null result \citep{ande88}, with a typical sensitivity $a/c\la (0.7-1.4)\times 10^{-18}$ s$^{-1}$ (2$\sigma$ upper limit on the gravitational acceleration of the spacecraft relative to the expected value; \citealt{ande95a, ande02}), about an order of magnitude less sensitive than the constraint based on the ephemerides of the outer planets. Recently \citet{ande02} have argued that {\it Pioneers} exhibit a constant anomalous acceleration of $a/c\simeq (2.9\pm 0.4)\times 10^{-18}$ s$^{-1}$, but this is probably due to non-gravitational effects.

Yet another constraint on the mass and the position of the companion can be derived based on the fairly accurate alignment of the spin of the sun and the orbital angular momenta of planets implying that there is no massive companion on a highly inclined orbit that would have forced the planetary orbits out of this configuration \citep{gold72, hogg91}. For a high-inclination companion at a large distance ($d_p\ga 100$ AU), this constraint is considerably more sensitive than the acceleration method.

A planetary companion to the solar system would reflect solar light and therefore could be detected in the optical or near-infrared by direct observations; furthermore, its own thermal radiation would be detectable in the mid- to far-infrared. Neither the optical surveys by \citet{tomb61} and \citet{kowa89}, nor the IRAS All-Sky Survey \citep{neug84} discovered any massive companions (other than Pluto). The optical surveys used a baseline of a few days to detect large-parallax objects, resulting in a typical distance sensitivity $d_p\la 400$ AU, whereas the IRAS survey was sensitive out to $400-2500$ AU, depending on the position on the sky. Neither survey covered 100\% of the sky, and the flux and parallax sensitivities varied across the sky in all three cases, as described in detail by \citet{kowa89} and \citet{hogg91}. In what follows we use Tombaugh's and IRAS surveys, as they had the largest sky coverage.

Modulo incompleteness, Tombaugh's and IRAS surveys had sufficient flux sensitivity to detect a rock-ice, low-albedo object with the mass given by equation (\ref{eq_conclusion}) out to about $d_p=90$ AU. For the calculation of the limiting distances for a rock-ice planet we followed \citet{hogg91} in assuming an albedo of 0.02, the limiting visual magnitude of Tombaugh's survey of 15.2, temperature of 280 K $(d_p/\mbox{1 AU})^{-1/2}$ (determined exclusively by solar heating), mass density of 2 g cm$^{-3}$ and the limiting flux of IRAS at 100\micron\ of 1 Jy. If the unseen companion is similar to giant planets rather than a rocky body, then its thermal radiation is dominated by its evolutionary cooling rather than by solar heating, and its radius is almost independent of mass. We used cooling models of brown dwarfs by \citet{alla01} which cover the range of masses from 1 $M_J$ to the hydrogen burning limit 0.08 $M_{\odot}$. In these models the temperature is a function of mass and age, and the isochrone at 5 Gyr can be approximated as 87 K $(m_p/M_J)^{0.57}$. A gas giant with mass given by equation (\ref{eq_conclusion}) would be detectable by IRAS in the 60 and 100\micron\ bands farther than about 70 AU out to the distances when the parallax of such object would be undetectable. The optical emission from a giant planet is reflected sunlight, with typical albedo of up to $0.5-0.6$, and is almost independent of its mass (insofar as the radius is independent of the mass). A giant planet with Jupiter's radius and albedo of 0.5 would be detectable in Tombaugh's survey out to 290 AU. 

To summarize the previous paragraphs, the acceleration method is more sensitive to distant ($d_p\ga 300$ AU) gas giants or rock-ice bodies than Tombaugh's survey, but it is not as sensitive as IRAS survey to a gas giant at $d_p\ga 70$ AU, out to about 400 AU, where IRAS parallax sensitivity becomes an issue. An alternative exotic possibility is that the companion is a dark compact object (a neutron star or a black hole) which could escape detectability in either survey. 

A mass given by equation (\ref{eq_conclusion}) may be detectable in a future high-precision astrometric survey through microlensing of background stars \citep{cowl83}, but would have been undetectable by Hipparcos. The minimum and the maximum masses detectable by this method are limited by the time coverage of the survey, since the microlensing event should be caught by at least one survey observation, but proceed faster than the total duration of the survey. The resulting range of masses that can be detected at distances $d_p\la 2500$ AU is 
\begin{equation}
m_p\in \left(\frac{1}{n}, 1\right) \times 60 M_J \times \left(\frac{\beta}{10\;\mu\mbox{as}}\right)\left(\frac{\tau}{\mbox{5 years}}\right),
\end{equation}
where $\beta$ is the astrometric sensitivity, $\tau$ is the total duration of the mission and $n\simeq 70$ is the typical number of survey scans of a given field. The numerical values of these parameters are given for the GAIA survey (to $V=15$ mag). A more detailed discussion of the sensitivity of GAIA to distant companions is given by \citet{gaud05}. 

\section*{Acknowledgments}

NLZ would like to thank Michael Strauss and Bohdan Paczy\'nski for useful discussions and Joseph Taylor and Edmund Bertschinger for comments regarding the gravitational wave background. The authors would like to thank Victoria Kaspi for very helpful comments and several initial discussions and the anonymous referee for suggestions. This research has made extensive use of the Australia National Telescope Facility pulsar database, http://www.atnf.csiro.au/research/pulsar/psrcat/. NLZ acknowledges the support of the Charlotte Elizabeth Procter Fellowship. ST acknowledges the support of NASA grant NNG04H44g.

\clearpage

\clearpage

\clearpage

\clearpage

\clearpage

\clearpage

\clearpage

\clearpage
\begin{deluxetable}{lcccccccccccccccclc}
\pagestyle{empty}
\rotate
\tabletypesize{\scriptsize}
\tablewidth{0pt}
\setlength{\tabcolsep}{0.03in}
\tablecaption{Constraints on the acceleration from pulsars in binary systems and pulsating white dwarfs}
\tablehead{ name & $\alpha$ & $\delta$ & $\dot{P}_{obs}/P$ & $\sigma_{obs}/P$ & $\dot{P}_{th}/P$ & $\sigma_{th}/P$ & $\dot{P}_{pm}/P$ & $\sigma_{pm}/P$ & $\dot{P}_{gal}/P$ & $\sigma_{gal}/P$ & $\bf \Delta\dot{P}/P$ & $\bf \sigma_{tot}/P$ & $\mu$, mas/yr & $d$, pc & $l$ & $b$ & comp. & ref.}
\startdata
PSR~J0437$-$4715 & 69.32 & -47.25 & 7.3 & 0.4 & -0.00073 & 0.00007 & 6.71 & 0.14 & -0.065 & 0.015 & {\bf 0.69} & {\bf 0.43} & 140.892(6) & 139(3) & 253.39 & -41.96 & WD & 1 \\
PSR~J1012+5307 & 153.14 & 53.12 & 0 & 1.9 & -0.21 & 0.05 & 1.3 & 0.2 & -0.11 & 0.04 & {\bf -1.1} & {\bf 1.9} & 25.3(4) & 840(90) & 160.35 & 50.86 & WD & 2 \\
PSR~B1534+12 & 234.29 & 11.93 & -3.55 & 0.36 & -5.28 & 0.02 & 1.10 & 0.33 & -0.18 & 0.07 & {\bf 0.81} & {\bf 0.49} & 25.5(5) & 700(200) & 19.85 & 48.34 & NS & 3 \\
PSR~J1713+0747 & 258.46 & 7.79 & 0 & 0.1 & -0.052 & -0.009 & 0.11 & 0.01 & -0.089 & 0.032 & {\bf 0.03} & {\bf 0.11} & 6.296(14) & 1100(100) & 28.75 & 25.22 & WD & 4 \\
PSR~B1855+09 & 284.40 & 9.72 & 0.6 & 1.2 & -0.00011 & 0.00002 & 0.085 & 0.026 & 0.0030 & 0.0010 & {\bf 0.5} & {\bf 1.2} & 6.16(11) & 900(300) & 42.29 & 3.06 & WD & 5 \\ 
PSR~B1913+16 & 288.87 & 16.11 & -86.66 & 0.03 & -86.0867 & 0.0007 & 0.097 & 0.027 & -0.54 & 0.12 & {\bf -0.13} & {\bf 0.12} & 2.6(3) & 5900(940) & 49.97 & 2.12 & NS & 6 \\
\hline
PSR~J0621+1002 & 95.34 & 10.04 & 0.0 & 7.0 & -0.0011 & 0.0003 & 0.058 & 0.017 & 0.12 & 0.04 & {\bf -0.2} & {\bf 7.0} & 3.5(6) & 1900(500) & 200.57 & -2.01 & WD & 7 \\
PSR~J0737$-$3039 & 114.46 & -30.66 & -136 & 9 & -141.21 & 0.18 & $<$0.16 & & -0.030 & 0.010 & {\bf 5.3} & {\bf 9.1} & $<$10.5 & 600(200) & 245.24 & -4.50 & 2PSR & 8 \\
PSR~J1141$-$6545 & 175.28 & -65.76 & -25 & 6 & -22 & 6 & $<$1.3 & $<$0.7 & -0.3(?) & 0.05(?) & {\bf -4} & {\bf 9} & 12(3) & $>$3800 & 295.79 & -3.86 & WD & 9 \\
PSR~J2145$-$0750 & 326.46 & -7.84 & 0.0 & 4.1 & -0.0009 & 0.0002 & 0.25 & 0.07 & -0.15 & 0.05 & {\bf -0.1} & {\bf 4.1} & 14.1(7) & 500(130) & 47.78 & -42.08 & WD & 10 \\
G117$-$B15A & 141.07 & 35.28 & 10.7 & 6.5 & 17.2 & 4.6 & 4.3 & 1.6 & -0.04 & 0.02 & {\bf -10.7} & {\bf 8.2} & 136(2) & 95(36) & 189.06 & 45.46 & none & 11
\enddata

\tablecomments{Right ascensions ($\alpha$) and declinations ($\delta$) are given in degrees for the epoch 2000; they are known with much better accuracy than given in the Table. All $\dot{P}/P$ and their uncertainties $\sigma/P$ are in units $10^{-18}$ s$^{-1}$. For pulsars in binary systems $P$ and $\dot{P}$ refer to their orbital periods and orbital period derivatives and for the DA white dwarf G117$-$B15A these are the pulsation period and its derivative. $\Delta\dot{P}\equiv\dot{P}_{obs}-\dot{P}_{th}-\dot{P}_{pm}-\dot{P}_{gal}$, $\sigma_{tot}^2\equiv \sigma_{obs}^2+\sigma_{th}^2+\sigma_{pm}^2+\sigma_{gal}^2$. For pulsars in binary systems, the error in the theoretical values is due to the error in the mass measurements, whereas for the white dwarf it is mostly due to the range of acceptable theoretical models, with mass being the primary source of uncertainty. Used in calculation of the relative Galactic acceleration and the proper motion correction are $\mu$ (proper motion in mas year$^{-1}$), $d$ (distance to the object in parsecs), $l$ and $b$ (Galactic longitude and latitude in degrees). Galactic acceleration corrections are calculated based on the P90 model for the potential of the Galaxy. The proper motion corrections and the Galactic acceleration corrections agree within the errors with those given in the literature, where available, except for the Galactic acceleration corrections for PSR~J2145$-$0750 \citep{lohm04} and PSR~J1713+0747 \citep{spla05} which are unimportant in both cases. The type of companion (WD -- white dwarf, NS -- neutron star, 2PSR -- the system consists of two pulsars) is listed in the second-to-last column. Above the horizontal line are `group 1' objects. }
\tablerefs{1 -- \citet{vans01}; 2 -- \citet{lang01}; 3 -- \citet{stai98}; 4 -- \citet{spla05}; 5 -- \citet{kasp94}; 6 -- \citet{damo91, weis03, weis04}; 7 -- \citet{spla02}; 8 -- \citet{lyne04, kram05}; 9 -- \citet{bail03}; 10 -- \citet{lohm04}; 11 -- \citet{brad98, kepl00}}
\end{deluxetable}

\clearpage
\begin{deluxetable}{ccccccc}
\pagestyle{empty}
\rotate
\tablewidth{0pt}
\tabletypesize{\footnotesize}
\setlength{\tabcolsep}{0.03in}
\tablecaption{Constraints on the acceleration of the solar system by different methods}
\tablehead{ method & $2\times 10^{-19}$ s$^{-1}$ & $5\times 10^{-19}$ s$^{-1}$ & $1\times 10^{-18}$ s$^{-1}$ & $2\times 10^{-18}$ s$^{-1}$ & $3\times 10^{-18}$ s$^{-1}$ & $4\times 10^{-18}$ s$^{-1}$}
\startdata
MSPs & 0\% & 3\% & 44\% & 99\% & 100\% & 100\% \\
PSR~B1913+16 & 21\% & 50\% & 75\% & 88\% & 92\% & 94\% \\
six pulsars in binary systems (group 1) & 25\% & 76\% & 94\% & 99.5\% & 100\% & 100\% \\
all eleven objects from Table 1 & 25\% & 76\% & 94\% & 99.5\% & 100\% & 100\% 
\enddata
\tablecomments{This table lists the fractions of the directions on the sky for which the acceleration of the solar system (given as $a_{\odot}/c$) can be ruled out at a given level with a 95\% confidence (see also Figure \ref{pic_sum}). For example, using MSPs accelerations $a_{\odot}/c>1\times 10^{-18}$ s$^{-1}$ can be ruled out for 44\% of the sky at 95\% confidence.}
\end{deluxetable}

\end{document}